\documentclass[aps,prl,reprint,superscriptaddress,preprintnumbers,floatfix]{revtex4-2}
\usepackage{amsmath,amssymb,amsfonts}
\usepackage{graphicx}
\usepackage{color}
\usepackage{mathtools}
\usepackage{bm}

\usepackage{siunitx}
\usepackage[bookmarksnumbered,raiselinks,breaklinks]{hyperref}
\hypersetup{
    colorlinks,%
    citecolor=blue,%
    filecolor=blue,%
    linkcolor=blue,%
    urlcolor=blue
}

\graphicspath{{./}}

\newcommand*{\real}{\ensuremath{ \mathbb{R} }}

\newcommand*{\defeq}{\coloneqq}
\newcommand*{\from}{\colon}
\newcommand*{\where}{\mid}

\newcommand*\bb[1]{\mathbf{#1}}
\DeclarePairedDelimiter\abs{\lvert}{\rvert}

\begin{document}

%\title{How to optimally invert a swing, and win at kiiking}
\title{Optimal strategies for kiiking: active pumping to invert a swing}
% Optimal control and learning for kiiking, an Estonian swinging sport
%\title{Optimal control and reinforcement learning strategies for the kiiking sport}

\author{Petur Bryde}
\affiliation{Paulson School of Engineering and Applied Sciences, Harvard University, Cambridge, MA 02138.}
\author{Ian C. Davenport}
\affiliation{Department of Physics, Harvard University, Cambridge, MA 02138.}
\author{L.\ Mahadevan}
\email{lmahadev@g.harvard.edu}
\affiliation{Paulson School of Engineering and Applied Sciences, Harvard University, Cambridge, MA 02138.}
\affiliation{Department of Physics, Harvard University, Cambridge, MA 02138.}
\affiliation{Department of Organismic and Evolutionary Biology, Harvard University, Cambridge, MA 02138.}

\begin{abstract}
Kiiking is an extreme sport in which athletes alternate between standing and squatting to pump a stationary swing till it is inverted and completes a rotation.  A minimal model of the sport may be cast in terms of the control of an actively driven pendulum of varying length to determine optimal strategies. We show that an optimal control perspective, subject to known biological constraints,  yields time-optimal control strategy similar to a greedy algorithm that aims to maximize the potential energy gain at the end of every cycle. A reinforcement learning algorithms with a simple reward is consistent with the optimal control strategy. When accounting for air drag, our theoretical framework is quantitatively consistent with experimental observations while pointing to the ultimate limits of kiiking performance.
    %
    % old
    %Pumping a swing is a paradigmatic example of a parametric oscillator, and used as an
    %analog for many problems from hydrodynamics to lasers. It is also the basis for an
    %extreme sport whereby athletes compete to pump a 5-7 m swing to make complete rotations
    %in the shortest time possible. To understand how, we use available experimental data,
    %constraints for human performance and examine the problem from two complementary
    %perspectives: optimal control and reinforcement learning. We find that ....
\end{abstract}

\maketitle
%\printinunitsof{pt}\prntlen{\textwidth} % 510pt
%\printinunitsof{pt}\prntlen{\linewidth} % 246pt
%
% ==========================================================================================
% Start of main text
% ==========================================================================================

% Intro notes

% Tea, Falk, 1968, Pumping on a swing
% Burns, 1970, More on pumping a swing
%   - Mathieu-Hill eq.
% Curry, 1976, How children swing
%   - viewing the motion as an example of parametric amplification
%   - in the linear regime, amplitude of the swing's motion increases exponentially
%     (Also the total energy increases exponentially)
% Arnold, 1978, Mathematical methods

% Parametric resonance: Burns, Curry, Arnold, Berry

%An analog effect can be observed for a mechanical pendulum, which can be pumped by
%periodically changing a parameter such as the pendulum length. This is well exemplified by a
%children’s swing, where the swinging child produces the pumping via periodically changing
%its center of mass (and thus effectively the length of the pendulum) twice every swing
%period [10]. Thus, the parametric pump in this case has a frequency twice the swinging
%frequency.

% Note that this is different from the usual way of pumping a playground swing from the
% standing position, by rotating the body back and forth.\cite{case1996}

% greedy strategy maximizes the amplitude after any given number of half-oscillations

% TODO captions
% \theta_n vs n. n = nth extreme displacement of the pendulum
% |\theta_n| = magnitude of the nth extreme displacement

\begin{figure}
    \centering
    \includegraphics[width=1.0\linewidth, trim={0.0cm 0.0cm 0.0cm 0.0cm}, clip]{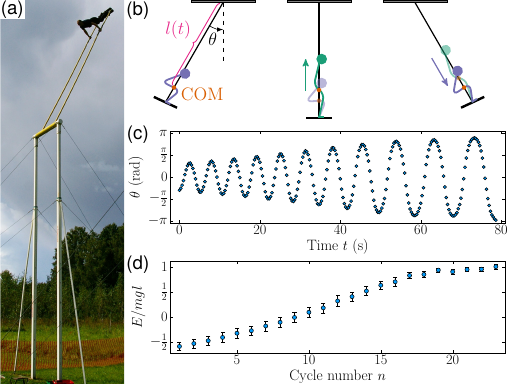}%
    \caption{ (a) An athlete on a tall kiiking swing. The arms of the swing are rigid, and
        the athlete is attached to the base by their feet.
        Adapted from Estonian Kiiking Association~\cite{kiikingphoto}, \copyright 2014 Estonian
        Kiiking Association.
        (b) A schematic of an athlete pumping the kiiking swing by squatting and standing.
        The swing-athlete system is modeled as a simple (point-mass) pendulum with variable
        length \(l(t)\). Over the course of one cycle, the athlete quickly stands up when
        the swing is near its lowest point, and squats down as the swing reaches its highest
        point. This stand-squat cycle is repeated twice over each oscillation of the swing's
        motion.
        (c) The angle \(\theta\) measured from experiments~\cite{kiikingvideoa} as a function
        of time for a successful kiiking attempt. The length of the swing is approximately
        \SI{7}{\meter}.
        (d) Nondimensionalized energy \(E/m g l_0 \approx -\cos \theta\) after the \(n\)th
        cycle, i.e.\ at the \(n\)th extremum of \(\theta\).
    }
    \label{fig:fig1}
\end{figure}

% bar, shaft, rod, pole
\paragraph{Introduction.}
Sports offer a fertile playground for the interaction of physics, physiology and cognitive neuroscience, raising questions about how humans learn and execute extreme motor tasks, sometimes accompanied by fame and fortune. The playground swing offers a humble and familiar example: a child wiggles on it randomly at first, but soon learns to move their body rhythmically, leading to limited amplitudes, but unlimited pleasure! But how does an individual learn to swing? What are the optimal strategies for pumping a swing? And how do physical and biological constraints enter in constraining the solution? Here, we study an extreme version of this problem termed Kiiking, invented in Estonia \cite{kiikingchannel}.  An athlete  strapped on a platform connected to an especially long ($\sim~7$ m) rigid swing (Estonian: \emph{kiik}) made of rigid bars Fig.~\ref{fig:fig1}(a) pumps the swing by standing and squatting, with the goal of inverting the longest possible swing and completing a full rotation within the shortest time.   
%old
%Kiiking is a national pastime and individual competitive sport invented in Estonia.
%In this sport, a specially designed swing (Estonian: \emph{kiik}) is used,
%where the seat is connected to the fulcrum by rigid bars of adjustable length,
%Fig.~\ref{fig:fig1}(a). The athlete stands fastened to the seat and pumps the swing by
%appropriately timed standing and squatting, with the goal of passing over the fulcrum,
%thereby completing a full rotation. The difficulty increases with the length of the swing
%bars, and athletes compete to achieve a full rotation on the longest swing possible.
%As a relatively new sport, there are many practical questions related to kiiking that
%have not been addressed, such as relative importance of timing and power,
%and what the optimal pumping strategy is.
%Another question of interest is how an athlete might learn the optimal strategy from
%experience.

Swings can be minimally modeled as an active pendulum, driven by either leaning back and forth ~\cite{case1990,case1996}, or by standing and squatting~\cite{wirkus1998}. In the linearized, small amplitude limit, the standing-squatting mode   Fig.~\ref{fig:fig1}(a), modeled as a pendulum with time-varying length \(l\), is a prototypical example of a parametrically driven oscillator~\cite{burns1970, berry2018}. %In this limit, an exponentially growing amplitude results if the length is periodically varied with a frequency which is approximately twice the natural frequency of the pendulum. 
However, to properly understand kiiking from a neurophysical perspective requires one to go beyond this and address the nonlinear problem from the perspective of optimal strategies for modulating the length, and further how this might be learned. A step in this direction takes the perspective of optimal control theory and  maximize the angle
\(\theta\) of the swing at its highest point over a half-period
e.g.~\cite{lavrovskii1993}, or minimize the time needed to reach a given
target angle or target potential
energy~\cite{piccoli1995time,kulkarni2003time,piccoli_kulkarni2005,luo1998time}.
Assuming that the length \(l\) can change discontinuously leads to the following intuitive
result: stand up at the lowest point of the swing, and squat at the highest point. However, biological and physical constraints limit the rate at which any athlete can stand and squat, especially at higher angular speeds when fighting gravity. Here we
combine the analysis of publicly available videos of kiiking~\cite{kiikingvideoa,kiikingvideob,kiikingvideoc,kiikingvideod,kiikingvideoe} to
extract time series, and use simple estimates of constraints on human athletic performance, e.g.\
maximum power exerted to constrain a minimal model of kiiking in terms of an extensible
controllable pendulum. 
\paragraph{Experimental data analysis.}
In order to develop a quantitative understanding of kiiking strategies we found videos of
kiiking online \cite{kiikingchannel,kiikingphoto} and took snapshots of the videos at a rate of 3 Hertz, which we 
analyzed using the Fiji software platform~\cite{fiji} to measure the angle the swing makes with the
vertical, as shown in Figure \ref{fig:fig1}(c) as the athlete makes a
full swing up to 180$^{\circ}$ (see SI SI). To estimate the speed and power limits on human performance, we note that athletes can squat/standup in about a second,
consistent with the maximum rate of standing from the video data of about $140$ cm/s,
and that the reported peak power output during a jump squat is on the order of 5000 W~\cite{cormie2007}.
We use these estimates later to set simulation parameters as we search for optimal swinging strategies.

\paragraph{Mathematical model.}
We model the kiiking system as a pendulum with a bob of mass \(m\)
connected to a pivot by a massless rigid rod of variable length \(l\), with $l_{-} < l < l_{+}$, the bounds corresponding to squatting and standing, and \(\theta\) the angle between the pendulum and the downward vertical direction Figure~\ref{fig:fig1}(b).
Neglecting any motion of the center of mass perpendicular to the swing arms, as well as
friction and air resistance for now, the kinetic and potential energy of the system are
\( T = \frac{1}{2} m \big[ (l \dot{\theta})^2 + \dot{l}^2 \big] \) and
\( V = - m g l \cos \theta\) respectively.
Defining the conjugate momentum \(p = m l^2 \dot{\theta}\), the
evolution of the state \(\bb{x} = (\theta, p, l)\) is given by Hamilton's equations
\begin{equation}
    \label{eom}
    \begin{aligned}
        \dot{\theta} &= \frac{1}{m l^2} p,\\
        \dot{p} &= -m g l \sin \theta,\\
        \dot{l} &= u,
    \end{aligned}
\end{equation}
where \(u\) is the rate of change of \(l\), which we take as the control variable.
We nondimensionalize the system~\eqref{eom} by choosing units so that \(m = l_0 = t_0 = 1\)
where \(l_0 = (l_{+} + l_{-})/2\) is a characteristic length scale and
\(t_0 = \sqrt{l_0/g}\) is a corresponding time scale.
We write the nondimensionalized equations in the form
\begin{equation}
    \label{eomaffine}
    \begin{gathered}
        \dot{\bb{x}} = \bb{f}(\bb{x}, u) = \bb{F}(\bb{x}) + \bb{G}(\bb{x}) u, \;
        \bb{x}(0) = \bb{x}_0,\\
        \bb{F}(\bb{x}) =
        \begin{pmatrix}
            p/l^2\\
            - l \sin \theta\\
            0
        \end{pmatrix}, \quad
        \bb{G}(\bb{x}) =
        \begin{pmatrix}
            0\\
            0\\
            1
        \end{pmatrix}
    \end{gathered}
\end{equation}
to emphasize that the system is affine in the control \(u\).
For simplicity, we only consider initial conditions of the form
\(\bb{x}_0 = (\theta_0, 0, l_{+})\) where \(\theta_0 \neq 0\).

Our goal is to find a control \(u\) so that the system reaches the target set
\(S = \{(\theta, p, l) \where \theta = \pm \pi\}\) in minimum time, subject to certain
constraints on \(u\) and the trajectory \(\bb{x}(t)\), which will be detailed below.
The control may be given in the form of an open-loop control \(u = u(t)\), or
else as a feedback control policy \(u = \pi(\bb{x})\).
Importantly, the bound \(l_{-} \leq l \leq l_{+}\) means that, for most initial
conditions, any control which steers the system to \(S\) must involve several \emph{cycles}
of squatting followed by standing. These individual cycles will be analyzed first before
turning to the full optimal control problem.

To gain some intuition about the system, we note that the rate of change of the (nondimensionalized) energy is
\begin{equation}
    \label{power}
    %\dot{E} = m u \dot{u} - m \big( l \dot{\theta}^2 + g \cos \theta\big) u.
    %\dot{E} = m u \dot{u} - ( p^2/(m l^3) + m g \cos \theta) u.
    \dot{E} = u \dot{u} - ( p^2/l^3 + \cos \theta) u.
\end{equation}
The first term vanishes over a cycle, so we consider only
the second term. For the system to gain energy, we must take \(u < 0\) (corresponding to
standing up) when the mass is near its lowest point, so that
\(p^2/l^3 + \cos \theta\) is maximal. Conversely, we should take \(u > 0\) near the
highest point in order to minimize energy losses during the squatting phase.
In our chosen units, the minimum energy needed to reach the target set \(S\) is simply
\(E = l_{-}\).

We make the natural assumption that the rate \(u = \dot{l}\) at which the length of the
pendulum can be changed is bounded by some maximum, \(\abs{u} \leq u_m\).
Motivated by equation~\eqref{power}, we also impose a \emph{power bound} of the form
%\(- m \big( l \dot{\theta}^2 + g \cos \theta\big) u \leq P_m\)
%\(-(p^2/(m l^3) + m g \cos \theta) u \leq P_m\)
\(-(p^2/l^3 + \cos \theta) u \leq P_m\)
%\begin{equation}
%    \label{powerbound}
%    g(\bb{x}, u) = - m \big( l \dot{\theta}^2 + g \cos \theta\big) u \leq P_m
%\end{equation}
for some \(P_m > 0\) \footnote{Note that this bound does not take into account the contribution
\(u \dot{u}\) to \(\dot{E}\). This is done for the sake of simplicity; however,
this term is usually small compared to the other two terms when the constraint
is active, except on very short times when \(u\) almost has a jump-like discontinuity.}.
The two constraints imply that \(u^{-} \leq u \leq u^{+}\) where
\begin{equation}
    \label{upm}
    u^{\pm} = u^{\pm}(\bb{x}) = \pm\min\big(u_m, \frac{\mp P_m}{p^2/l^3 + \cos \theta}\big)
\end{equation}
Finally, we define the dimensionless parameter \(\Delta l \defeq (l_{+}-l_{-})/2\)
so that \(l\) is bounded between \(1-\Delta l\) and \(1 + \Delta l\).
For kiiking athletes, typical ranges for the dimensionless parameters are
\(0.04 \leq \Delta l \leq 0.08\),
\(0.1 \leq u_m \leq 0.15\), and \(0.1 \leq P_m \leq 0.25\).
%We remark that in the limit \(P_m \to \infty\), \(u_m \to \infty\), we obtain the problem considered in~\cite{luo1998time,piccoli_kulkarni2005}.
\begin{figure}
    \centering
    \includegraphics[width=1.0\linewidth, trim={0.0cm 0.0cm 0.0cm 0.0cm}, clip]{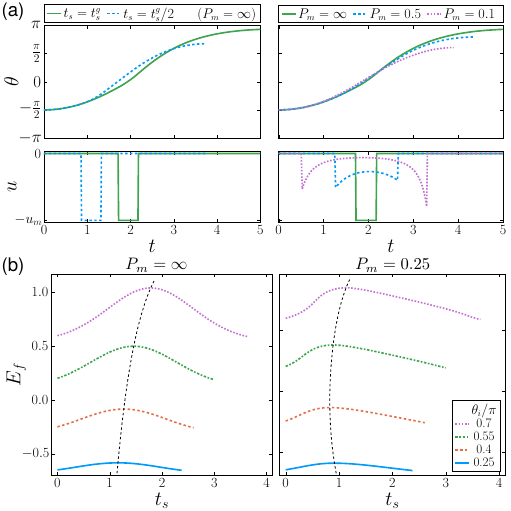}%
    \caption{Analysis of the standing phase of a stand-squat cycle. Starting at the
        switching time \(t = t_s\), the athlete decreases the swing length \(l\) at the
        maximum rate \(u = u^{-}\) (Eq.~\eqref{upm}) until \(l\) reaches the minimum
        value \(l_{-}\).
        (a) The angle \(\theta\) and control \(u = \dot{l}\) as functions of
        (nondimensionalized) time.
        Left: The switching time \(t_s\) is varied in the absence of power constraints
        (\(P_m = \infty\)). The greedy switching time \(t_s^{g}\) (solid green lines)
        maximizes the energy gained by the system.
        %Blue lines shows a sub-optimal switching time for comparison.
        Right: Solutions with \(t_s = t_s^{g}\) and varying power bounds \(P_m\).
        In both cases, \(\Delta l = 0.115\), \(u_m = 0.5\).
        (b) Energy \(E_f\) at the end of a cycle as a function of the switching time \(t_s\).
        Each solid curve corresponds to a different initial angle \(\theta_i = \theta(0)\).
        The dashed black curve indicates the greedy switching time \(t_s^{g}\)
        for each \(\theta_i\). For sufficiently large \(P_m\), \(t_s^{g}\) increases
        monotonically with \(\theta_i\).
        Here, \(\Delta l = 0.05\), \(u_m = 0.1\).
    }
    \label{fig:fig2}
\end{figure}
\paragraph{Greedy control algorithm.}
%***
%Figure~\ref{fig:fig2} shows numerical simulation results obtained using an optimal control package \cite{Casadi}
%for the solution of the optimal control problem
%${\rm Argmin}_u T$ to reach the target set $S$ subject to the constraints given by~\eqref{eom},~\eqref{upm}.
%***
Figure~\ref{fig:fig2} shows numerical solutions of equation~\eqref{eomaffine} subject to the constraints \(l_{-} \leq l \leq l_{+}\) and
\(u^{-} \leq u \leq u^{+}\) (see Eq.~\ref{upm}) for the standing phase of a single cycle.
At \(t=0\), the swing starts from rest at an initial angle \(\theta_i\) and remains
in the lowest position (\(l = l_{+}\)) until a \emph{switching time} \(t_s\), when \(u\)
is taken to be the minimum value allowed by the rate and power constraints, i.e.\
\(u(t) = u^{-}\) where \(u^{-}\) is defined in equation~\eqref{upm}.
This \(u(t)\) is maintained until \(l\) reaches the minimum length \(l_{-}\), after which
\(u(t)\) is identically zero. In particular, \(u(t)\) is piecewise constant in the absence
of power constraints (\(P_m = \infty\)).
Figure~\ref{fig:fig2}~(a) shows some representative solutions and the corresponding
controls.
The angle \(\theta\) at the end of the standing phase, and thus the energy gained by the
system, is seen to depend sensitively on the switching time \(t_s\).
The dependence of the final energy \(E_f\) on the switching time is shown explicitly in
Figure~\ref{fig:fig2}~(b).
We denote by \(t_s^{g}\) the \emph{greedy switching time}, which maximizes the
energy at the end of the standing phase.
In the limiting cases \(P_m \to \infty\), \(u_m \to \infty\),
the greedy switching time is precisely when \(\theta = 0\), i.e.\ when the swing reaches
its lowest point. This is in agreement with previous results~\cite{luo1998time,piccoli_kulkarni2005}.
Since the period of a pendulum increases with amplitude,
\(t_s^{g}\) increases monotonically with \(\theta_i\) for sufficiently large \(P_m\).
In the power-constrained case, the dependence is more complicated due to a competing effect:
As \(\theta_i\) increases, the power constraint becomes more strict,
leading to smaller \(\abs{u}\) and earlier switching times \(t_s^{g}\).

The preceding discussion suggests a \emph{greedy} control strategy, obtained by
repeatedly switching between standing and squatting at the greedy switching times.
Specifically, let \(u(t) = u^{(i)}(t)\) for \(t^{(i)} \leq t < t^{(i+1)}\) (\(i = 1,\ldots, N\))
where the \(u^{(i)}\) vary cyclically between the values \(u = u^{-}\), \(u = 0\) and
\(u = u^{+}\), and the switching times \(t^{(i)}\) are chosen greedily, i.e.\ in order to
maximize the energy at the end of each stand-squat cycle. By construction, this strategy
achieves the task of reaching \(S\) in the fewest number of stand-squat cycles.
Moreover, it approximates the behavior observed from kiiking athletes.
As we will show, the time-optimal solution has the same structure as the greedy strategy,
but with slightly different switching times.
We will now precisely pose the time-optimal control problem.

\paragraph{Time-optimal control.}
We say that a measurable function \(u \from \real_{+} \to U\), \(U = [-u_m, u_m]\), is an
\emph{admissible} control if \(u\) and the corresponding state trajectory \(\bb{x}\), solution
of~\eqref{eom}, satisfy the inequality constraints
\begin{equation}
    \label{stateconstraints}
    g(\bb{x}(t), u(t)) \geq 0, \quad
    \bb{h}(\bb{x}(t)) \geq 0,
\end{equation}
where
\begin{equation*}
    g(\bb{x}, u) = P_m + (p^2/l^3 + \cos \theta) u, \;
    \bb{h}(\bb{x}) =
    \begin{pmatrix}
        l_{+} - l\\
        l - l_{-}
    \end{pmatrix}.
\end{equation*}
Define the terminal time \(t_f = t_f(u)\) as the first time that the corresponding
trajectory \(\bb{x}\) hits the target set
\(S = \{(\theta, p, l) \where \theta = \pm \pi\}\). Then the time-optimal control problem
(TOCP) for the kiiking system can be stated as follows: Given an initial state \(\bb{x}_0\),
find an admissible control \(u \in \mathcal{U}\) which minimizes \(t_f\) subject to the
state constraints~\eqref{stateconstraints}.

The presence of the pure state constraints, represented by \(\bb{h}(\bb{x})\) in
equation~\eqref{stateconstraints}, makes this a difficult problem to solve by variational
methods.
A set of necessary conditions for a control-trajectory pair to solve the optimal control
problem are provided by the Pontryagin Maximum Principle (PMP) for both mixed control-state
inequality constraints as well as pure state constraints~\cite{sethi2021}. Defining the control Hamiltonian associated with the TOCP as
\begin{equation*}
    H(\bb{x}, \bm{\lambda}, u) = \bm{\lambda} \cdot (\bb{F}(\bb{x}) + u \bb{G}(\bb{x})),
\end{equation*}
the time derivative of the constraint function \(\bb{h}\) is
\begin{equation*}
    \bb{h}^{1}(\bb{x}, u) = \nabla \bb{h}(\bb{x}) \cdot \bb{f}(\bb{x}, u) = (-u, u)^{T}
\end{equation*}
and we define the restricted control set
\begin{equation}
    \label{restrictedcontrolset}
    \begin{gathered}
        \tilde{U}(\bb{x}) \defeq \{u \in U \where
            g(\bb{x}, u) \geq 0 \text{ and }\\
            h^{1}_i(\bb{x}, u) \geq 0 \text{ if } h_i(\bb{x}) = 0, \; i=1,2
        \}.
\end{gathered}
\end{equation}
The PMP states that if \(u^{*} \from [0, t_f] \to U\) is an optimal control and
\(\bb{x}^{*}\) is the corresponding trajectory, then there exists a nowhere vanishing
function \(\bm{\lambda}^{*} \from [0, t_f] \to \real^3\) (the \emph{costate})
so that at any time \(t\) the function
\(u \mapsto H(\bb{x}^{*}(t), \bm{\lambda}^{*}(t), u)\) attains its maximum on the restricted
control region \(\tilde{U}(\bb{x}^{*}(t))\) at \(u = u^{*}(t)\).

In our case, the control Hamiltonian \(
H(\bb{x}, \bm(\lambda), u) = \bm{\lambda} \cdot \bb{F}(\bb{x}) + u \lambda_3
\) depends on \(u\) only through the term \(u \lambda_3\). The coefficient
\(\varphi(t) = \lambda_3(t)\) multiplying \(u\)
is called the \emph{switching function} and determines the structure of the optimal control.
On \emph{interior arcs}, i.e.\ at times \(t\) when the constraint \(\bb{h}\) is inactive,
the PMP states that
\(u^{*}(t) = u^{+}\) if \(\varphi(t) > 0\) and
\(u^{*}(t) = u^{-}\) if \(\varphi(t) < 0\).
In other words, \(u^{*}\) changes between the maximum and minimum value at
\emph{switching times} when \(\varphi\) switches sign.
This completely determines \(u^{*}(t)\) on interior arcs unless
a \emph{singular arc} occurs, i.e.\ if \(\varphi(t)\) vanishes on a nontrivial interval.
We find that singular arcs do not appear in our problem, so we do not consider them further.
A time interval \([t_1, t_2]\) on which the constraint \(\bb{h}\) is active is called
a \emph{boundary arc} and its endpoints \(t_1, t_2\) are called \emph{junction times}.
We see that in the interior of a boundary arc, we necessarily have \(u^{*}(t) = 0\).

\begin{figure}
    \centering
    \includegraphics[width=1.0\linewidth, trim={0.0cm 0.0cm 0.0cm 0.0cm}, clip]{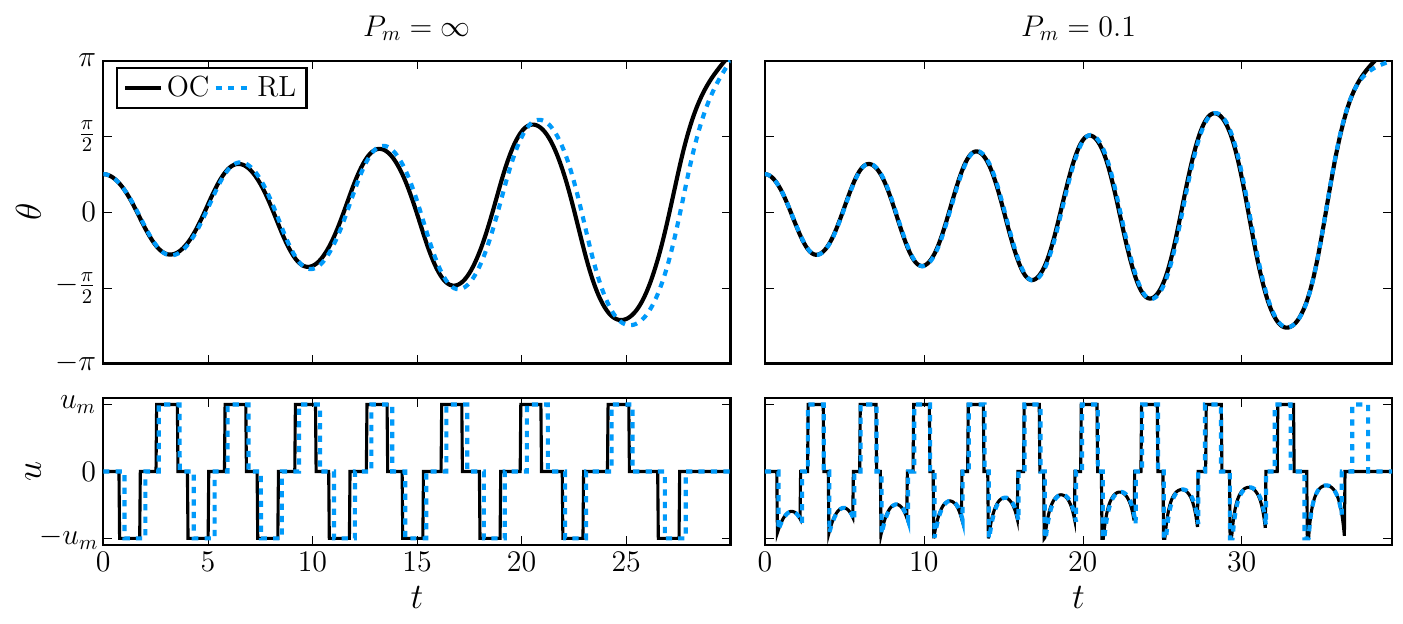}%
    \caption{Time-optimal solution of the kiiking problem.
        The angle \(\theta\) and control \(u = \dot{l}\) as functions of
        (nondimensionalized) time for the numerically computed optimal control solution (OC,
        black lines) and the solutions obtained from reinforcement learning (RL, blue
        lines).
        Left: When the system is not power-constrained, the OC solution consists
        of several interior arcs where \(u = \pm u_m\), separated by boundary arcs where
        \(\abs{l-1} = \Delta l\) and \(u = 0\).
        Right: In the presence of power constraints, \(u\) takes the value
        \(u^{\pm}\) (equation~\eqref{upm}) on the interior arcs.
        In both cases, the RL solution has the same form but with slightly different
        switching times. For both simulations, \(\theta_0 = \pi/4\), \(\Delta l = 0.05\),
        \(u_m = 0.1\).
        \label{fig:fig3}
    }
\end{figure}

Computing the costate \(\bm{\lambda}\) (and thus the switching function) requires
solving a difficult nonlinear multi-point boundary value problem. Alternately, we can use standard nonlinear
programming methods to determine \(u^{*}(t)\) in terms of the sequence of switching times and junction times after we
transcribe the time-optimal control problem into a finite-dimensional optimization problem~\cite{maurer2009}.
This approach requires an initial guess which is sufficiently close to the optimal switching times.
Since the greedy algorithm described in the previous section is near-optimal, it provides a suitable initial guess.
The computed optimal control was verified using a direct collocation method implemented in CasADi~\cite{Andersson2019} (See SI SIII).
The optimal control and corresponding \(\theta(t)\) are shown in Figure~\ref{fig:fig3} (black lines).
As expected, the optimal control has the same structure as the greedy strategy, and differs only slightly in the switching and junction times.

\paragraph{Reinforcement learning.}
Reinforcement learning (RL) provides an alternate approach to optimal decision and control problems \cite{sutton_barto}.  While RL methods do not provide the same optimality guarantees as control theory methods, they are nevertheless very powerful as direct approaches, but need to phrased in terms of a state
space $\mathcal{S}$, an action space $\mathcal{A}$, and a reward signal $\mathcal{R}$. For kiiking following the state space is given by Eq. \ref{eom}, $S_t = (\theta_t, p_t, l_t)$. Since the athletes must stand or squat as quickly as possible and otherwise do nothing, the action space is then $\mathcal{A} = \{-u_{m}, 0, u_{m}\}$, where $u_{m}$ is the
maximum rate the athlete can stand or squat. At every time step the RL agent chooses an
action and then follows the dynamics in equation \ref{eom}, while ensuring that the power bounds are not
exceeded at each time step.

% \footnote{We pre train the agent with a reward that includes the swing energy at
% each time. The energy based reward is less sparse than the pure time based reward and helps
% the agent learn a near optimal policy. After pretraining we train the agent with a purely
% time based reward.}

To achieve the goal of training the agent to swing up to $\theta = \pi$ as quickly as possible, we supplement the time based reward with a reward which encourages incremental progress of the form
\begin{equation}
    R_t =
    \begin{cases}
        1 & \text{if } \theta = \pi\\
        -1 + E_t/E_{max} & \mathrm{otherwise}
    \end{cases}
\end{equation}
where $E_{max}$ is the gravitational potential energy of the swing at $\theta = \pi$ and
$E_t$ is the energy at time $t$. Energy increases as the swing cycles become larger in
amplitude so an energy based reward is similar, but not identical, to a time based reward
(see Fig \ref{fig:fig2}). While using informative intermediate rewards via
reward shaping  is frequently used in RL problems~\cite{ng1999policy}, we found no improvement in performance after starting with the hybrid time-energy optimal agent and retraining it with the time based reward. As such we
present results using the hybrid reward.
% We chose the reward to encourage the agent to swing up to $\theta=\pi$ as quickly as possible. For every time step the agent has not reached it's goal it gets penalized. If it finally does reach the goal, it gets a positive reward. See the appendix for more information on the training procedure.
% \begin{equation}
%     R_t =
%     \begin{cases}
%         1 & \text{if } \theta = \pi\\
%         -1 & \mathrm{otherwise}
%     \end{cases}
% \end{equation}
% After specifying the state space, action space, and reward we need to find a policy $\Pi: \mathcal{S} \rightarrow \mathcal{A}$ which maximizes the reward. Given some parameterized policy $\Pi = \Pi_\alpha$, policy gradient algorithms use gradient ascent to move the parameters in the direction which maximizes the reward. In other words we compute how the reward changes with respect to the parameters $\alpha$ and update $\alpha$ so that the reward increases.
Practically, we parameterized the policy as a neural network and use the Proximal Policy
Optimization (PPO)~\cite{ppo} algorithm, a variant of traditional policy optimization, to
update the network weights towards the optimal policy, and used the PPO implementation from
Stable Baselines 3~\cite{stable-baselines3}, a thoroughly tested software package with
implementations of various reinforcement learning algorithms (see SI SIII).
% We chose to use the classic Deep Q Network (DQN) learning algorithm \cite{dqn}. Instead of
% learning a policy directly DQN learns a function $Q = Q_\Pi(s, a)$ called the state-action
% value function. The function $Q_\Pi(s, a)$ describes the expected future returns given
% that the system is in state $s$, takes action $a$, and follows policy $\Pi$ thereafter. In
% DQN we approximate Q as a deep neural network and use Q learning techniques \cite{dqn,
% sutton_barto} to update the network parameters towards those of the $Q$ function for the
% optimal policy. Once we have the $Q$ function for the optimal policy $\Pi^*$, the action
% $a$ which maximizes $Q^*(s, a)$ determines which action to choose in a given state $s$.

% \begin{equation}
%     \Pi^*(s) = \underset{a}{\mathrm{argmax}} \hspace{1mm} Q^*(s, a)
% \end{equation}

% We used an existing DQN implementation in the Julia programming language.

Figure~\ref{fig:fig3} compares solutions obtained by the optimal control (OC)-based method
to solutions found by the reinforcement learning (RL) algorithm (see also Videos 1-2 in the
Supplemental Information). The solutions are identical apart from minor differences in
switching and junction times. In particular, both solutions agree qualitatively with the
general strategy of kiiking athletes, namely standing up as quickly as possible as
\(\theta\) passes through \(0\) and squatting near zeros of \(p\) (the extrema of
\(\theta\)).

\begin{figure}
    \centering
    \includegraphics[width=1.0\linewidth, trim={0.0cm 0.0cm 0.0cm 0.0cm}, clip]{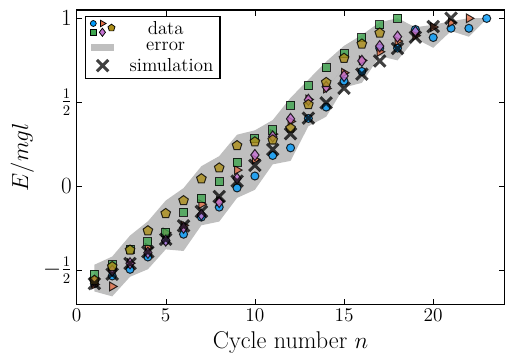}
    \caption{
        Nondimensionalized energy \(E/m g l\) at the end of each stand-squat cycle for five
        trials by different kiiking trials. The gray ribbon shows the envelope of the error
        bars for all five trials. In the two longest trials (circles and triangles)
        the length of the swing arms is approximately \SI{7}{\meter}. For the other trials,
        the length is smaller but not precisely known.
        Two of the series (diamonds and pentagons) show failed attempts, which ended without
        reaching \(\theta = \pi\).
        A simulation with parameters \(\Delta l = 0.05\), \(u_m = 0.125\), \(P_m = 0.125\),
        \(d = \num{2.75e-2}\) is shown for comparison, and captures the sigmoidal shape
        of the \(E(n)\) curves.
        \label{fig:fig4}
    }
\end{figure}

\paragraph{Comparison with data.}
To validate our model, we compare the computed OC solutions to experimental data collected from
trials by five different kiiking athletes~\cite{kiikingvideoa,kiikingvideob,kiikingvideoc,kiikingvideod,kiikingvideoe}.
In the absence of (mass and geometric) data on the athletes, it is not possible to directly compare \(\theta(t)\) time series with our predictions, and so we compare the maximum scaled potential energy
\(E(n)\) at the end of each stand-squat cycle \(n\), \(E = -\cos \theta\). In
figure~\ref{fig:fig4} we see that for the longer trials, the \(E(n)\) curves have a distinctive sigmoidal shape.
Recalling equation~\eqref{power}, we see that the energy gain is initially limited by the
small term \((p^2/l^3 + \cos \theta) u \approx p^2 u\), but at larger velocities, air resistance  becomes
non-negligible, and the \(E(n)\) curve levels off. To capture these qualitative features, we modify our model Eq.~\eqref{eom} by adding a quadratic drag term $   \dot{p} = - l \sin \theta - d \abs{p} p/l,$
where \(d\) is a dimensionless constant. Figure~\ref{fig:fig4} shows the corresponding \(E(n)\) using this form of the energy, for a challenging kiiking exercise requiring over twenty stand-squat cycles and captures the sigmoidal shape seen in data.

We note that adding the effects of air drag also has a qualitative implication for kiiking, i.e. for
some parameter values, the set \(S\) cannot be reached in finite time. To see this, we note that
for given \(u_m, \Delta l, d > 0\) there is, intuitively, a minimum
power \(P_m\) needed to complete the kiiking task. Furthermore, there is also a maximum swing length (minimum \(\Delta l\)) set by \(u_m, d\) above which the problem is infeasible for every \(P_m > 0\).
In other words, there is a theoretical maximum swing length that a given athlete can use to
successfully complete the kiiking task, regardless of the maximum power (see SI SIV).

%Figure \ref{fig:fig3} shows the results of using the two optimization methods to find
%swinging strategies. From watching videos of Kiikers we know that the general strategy is to
%squat at the peak of a swing cycle and stand up as the swing passes through $\theta = 0$.
%The optimal solutions agree with this intuition. In figure \ref{fig:fig3}a we see the agent
%goes from squatting as quickly as possible (positive $u$) near extrema in $\theta$ to
%standing as quickly as possible as $\theta$ passes through 0. Figure \ref{fig:fig3}b
%demonstrates the same phenomenon except there we have added the power bound. Instead of
%being able to stand up at $u_m$ we see the curved control profile in \ref{fig:fig3}b. We
%also see that the reinforcement learning solution does not quite match the known optimal
%strategy derived from control theory. In both cases the task is solved in the same number of
%swing cycles but the RL is slightly slower in each case. Nonetheless, in both the power
%bounded and unbounded case the feedback based RL solution is close to the open loop control
%theory solution.
%
%Finally we compare the optimal solution to experiment. Figure \ref{fig:fig4} shows the
%energy over time for the optimal control solution compared with experiments. Note we have
%added quadratic drag to make the simulation more realistic. The energy growth by cycle
%matches well between theory and experiment and indicates that real Kiikers are swinging in a
%near optimal manner.

% one advantage of the RL solution is that it is given in feedback form
\paragraph{Discussion.}
Inspired by the humble swing and the extreme Estonian sport of kiiking, we considered the dynamics of an actively pumped pendulum. Using various approaches derived from control and learning theory, we computed strategies for the feedforward (open-loop) control of the swing and how an athlete or a robot might learn to complete the kiiking task from repeated attempts. Our results are consistent with observations of experimental data and serve to highlight the role of explicitly including physiological limitations on dynamics and energetics. Beyond the specific study, our work points to how even seemingly simple problems in physics. e.g. the playground swing, can be a rich source of new questions that link physics, physiology and neuroscience, when approached from a constrained learning perspective.

\paragraph{Acknowledgments}
    We thank Ants Tamme for valuable discussion and information about kiiking, and the
    Estonian Kiiking Association for the use of their images and videos. We also acknowledge partial financial support from the Simons Foundation and the Henri Seydoux Fund.

P.B.\  and I.C.D.\ contributed equally to this work.

\end{document}